\documentstyle[editedvolume,numreferences,epsf]{crckapb}

\def\lbiblabel#1{$#1$} %
\def\lbibitem[#1]#2{\item[\lbiblabel{#1}]\if@filesw
{\def\protect##1{\string ##1\space}\immediate
\write\@auxout{\string\bibcite{#2}{#1}}}\fi\ignorespaces}

\def\ealla#1{{\rm e}^{#1}}
\def\O#1{{\cal O}\left (#1 \right )}
\def\half{{1\over 2}}
\def\eqref#1{(\ref{eq:#1})}
\def\llangle{\left\langle \negthinspace \left\langle}
\def\rrangle{\right\rangle \negthinspace \right\rangle}
\def\G{{\cal G}} \def\H{{\cal H}} \def\E{{\cal E}} \def\Z{{\cal Z}}
\def\PP{\mbox{\boldmath $\cal P$}}\def\DD{\mbox{\boldmath $\cal D$}}
\def\QQ{\mbox{\boldmath $\cal Q$}}\def\SS{\mbox{\boldmath $\cal S$}}
\def\W{{\cal W}}\def\A{{\cal A}}

\begin{opening}

\title{REPTATION QUANTUM MONTE CARLO}
\subtitle{a round-trip tour from classical diffusion to quantum mechanics}

\author{STEFANO BARONI$^*$}
\author{SAVERIO MORONI$^{\dagger}$}
\institute{SISSA -- Scuola Internazionale Superiore
di Studi Avanzati, \\ INFM -- Istituto Nazionale per la
Fisica della Materia, \\ via Beirut 2-4, 34014 Trieste Grignano, Italy}

\end{opening}
\runningtitle{REPTATION QUANTUM MONTE CARLO}

\makeindex

\begin{document}


\begin{opening}
\author{STEFANO BARONI AND SAVERIO MORONI}
\end{opening}

\begin{abstract} We present an elementary and self-contained account of
the analogies existing between classical diffusion and the
imaginary-time evolution of quantum systems. These analogies are used
to develop a new quantum simulation method which allows to calculate
the ground-state expectation values of local observables without any
mixed estimates nor population-control bias, as well as static and
dynamic (in imaginary time) response functions. This method, which we
name {\it Reptation Quantum Monte Carlo}, is demonstrated with a few
case applications to $^4$He, including the calculation of total and
potential energies, static and imaginary-time dependent density response
functions, and low-lying excitation energies. Finally, we discuss the
relations of our technique with other simulation
schemes. \end{abstract}



\section{Introduction}\label{sec:intro} The theory of stochastic processes
plays an important role in modern developments of quantum mechanics
both as a deep---and possibly not yet fully exploited---conceptual
method \cite{Nelson,Parisi} and as a powerful practical tool for the
computer simulation of interacting quantum systems \cite{QMC}. Quantum
Monte Carlo simulations mainly rely on the {\it static} properties of
one kind or the other of random walk that is used to sample the
ground-state wave-function or finite-temperature density matrix of a
system. Comparatively minor attention has been paid so far to the {\it
dynamical} properties of the random walks used in quantum
simulations. The main interest in these properties stems from the
study of excitation energies, a notoriously difficult and
ill-conditioned problem. These dynamical properties, also determine
the magnitude of autocorrelation times which are necessary to estimate
statistical errors.

The purpose of these lectures is to provide an elementary and
self-contained presentation of the deep relations existing between
diffusion in classical systems and the imaginary-time evolution of
quantum systems, and to develop some ideas based on these relations
\cite{caffarel} which lead to a new, promising technique for
performing quantum simulations. For reasons which will become apparent
in the following, we name this technique {\it Reptation Quantum Monte
Carlo} (RQMC). The main features of RQMC are that it is based on a
{\it purely diffusive} process without branching, that ground-state
expectation values of local observables can be evaluated without any
mixed estimates, and that static and dynamic (in imaginary time)
response functions are natural by-products of the ground-state
simulation. Although some of the ideas presented here are not new
\cite{caffarel}, our method does not suffer from the drawbacks which
affected previous their implementations.

\section{From classical diffusion to quantum mechanics} \label{sec:cl-qu}
The simplest phenomenological description of classical diffusion is
given by the Langevin equation: \begin{equation} dx = f(x) d\tau +
d\xi, \label{eq:langevin} \end{equation} where $x$ is some generalized
coordinate describing our system, $f(x) = -{\partial v(x) \over
\partial x}$ is the force acting on it---which we suppose to be
derivable from a potential---$\tau$ is time and $\xi(\tau)$ is a
Wiener process: $\langle d\xi\rangle = 0$; $\langle (d\xi)^2
\rangle=2d\tau$. Although the random walk generated by the Langevin
equation, Eq. \eqref{langevin}, can be given a mathematically
unambiguous meaning in the continuous limit, it is simpler---and more
useful in view of applications to quantum simulations---to specialize
to a given discretization of time: $\tau_k = k\times\epsilon$. The
corresponding random walk is then described by the discrete Markov
chain: \begin{equation} x_{k+1} = x_k + \epsilon f(x_k) + \xi_k,
\label{eq:discr-langevin} \end{equation} where the $\xi$'s are
uncorrelated Gaussian random variables of zero mean, $\langle \xi_i
\rangle =0$, and variance $2\epsilon$, $\langle \xi_i\xi_j \rangle =
2\epsilon \delta_{ij}$.

\subsection{The Fokker-Plank equation}\label{subs:fp} The time evolution
of the probability distribution for the variable $x$ is given by
the master equation which, in the one-dimensional case, reads (the
generalization to many dimensions is straightforward): \begin{equation}
P(x,\tau+\epsilon) = \int {\cal W}_\epsilon(x,y) P(y,\tau)~dy,
\label{eq:master} \end{equation} where \begin{equation} {\cal
W}_\epsilon(x,y) = {1\over \sqrt{4\pi\epsilon}} \ealla{-{(x-y-\epsilon
f(y))^2\over 4\epsilon}} \label{eq:tr-prob} \end{equation} is the
conditional probability that the system is in configuration $x$ at
time $\tau+\epsilon$, given that it is found in configuration $y$ at
time $\tau$. When the conditional probability is independent of $\tau$
(as it is in the present case) the corresponding Markov process is
said to be {\it homogeneous}. In order to convert the master equation,
Eq. \eqref{master}, into a differential equation, we perform a Taylor
expansion of its right-hand side in powers of the time step
$\epsilon$: \begin{equation} P(x,\tau+\epsilon) = {1\over\sqrt{4\pi
\epsilon}} \int \delta(x-y-\epsilon f(y)-\xi) \ealla{-{\xi^2\over
4\epsilon}} P(y,\tau) ~d\xi ~dy. \label{eq:sega} \end{equation} Taking
into account that $\xi \sim \sqrt{\epsilon}$, we now formally
Taylor-expand the $\delta$ function in powers of $(\epsilon f(y) +
\xi)$, and we obtain: \begin{equation} P(x,\tau+\epsilon) = \int
P(y,\tau) \left ( \sum_n {(-1)^n \over n!} ~\delta^{(n)}(x-y) \times
\left \langle (\epsilon f(y) + \xi )^n \right \rangle \right )~dy,
\label{eq:sega-2} \end{equation} where $\langle\cdot\rangle$ indicates
the Gaussian average of the polynomials in $\xi$, and $\delta^{(n)}$ is
the $n$-th derivative of the $\delta$ function: \begin{eqnarray} \left
\langle (\epsilon f(y) + \xi )^n \right \rangle &\equiv&
{1\over\sqrt{4\pi \epsilon}} \int_{-\infty}^\infty (\epsilon f(y) +
\xi )^n \ealla{-{ \xi^2\over 4\epsilon}} d\xi \nonumber \\ &=&
(-i\sqrt{\epsilon})^n H_n\left ( i{\sqrt{\epsilon}\over 2} f(y) \right
), \end{eqnarray} $H_n$ being the Hermite polynomial of order $n$
\cite{hermite}. To linear order in $\epsilon$, the first few values of
these Gaussian integrals are: $\langle \epsilon f(y) + \xi \rangle =
\epsilon f(y)$, $\langle (\epsilon f(y) + \xi )^2\rangle = 2\epsilon
+\O{\epsilon^2}$, $\langle (\epsilon f(y) + \xi )^3\rangle =
\O{\epsilon^2}$. Integrals of the derivatives of the $\delta$ function
are given by: $\int g(y)\delta^{(n)}(x-y) dy = g^{(n)}(x)$. Using this
relation, Eq. \eqref{sega-2} can be recast as: \begin{equation}
P(x,\tau+\epsilon) = P(x,\tau) + \epsilon \left ( -{\partial\over
\partial x} \left ( f(x) P(x,\tau)\phantom{^2}
\negthinspace \negthinspace \right ) + {\partial^2 P(x,\tau)
\over \partial x^2} \right ) + \O{\epsilon^2}. \label{eq:discr-FP}
\end{equation} Eq. \eqref{discr-FP} is the discrete-time version of
the Fokker-Planck \cite{stochastic-processes} equation which, in the
continuous limit, reads: \begin{equation} {\partial P(x,\tau) \over
\partial \tau} = {\partial^2 P(x,\tau) \over \partial x^2} - {\partial
\over \partial x} \left (f(x)P(x,\tau)\phantom{^2} \negthinspace
\negthinspace \right ). \label{eq:fokker-planck}\end{equation} The
fact that this equation is first order in time is strictly related to
the Markovian character of the random walk, Eq.
\eqref{discr-langevin}. It is easily verified that $P_s(x) \propto
\ealla{-v(x)}$ is a stationary solution of the Fokker-Planck equation,
Eq. \eqref{fokker-planck}. Using Eq. (\ref{eq:discr-FP}) one sees that
the stationary distribution of the {\it discrete} random walk,
Eq. \eqref{discr-langevin}, differs from $P_s$ by terms of order
$\epsilon$.

The conditional probability, ${\cal W}_\epsilon$, defined in
Eq. \eqref{tr-prob} can be seen as the {\it exact} probability that
the system described by the {\it discrete} Markov chain,
Eq. \eqref{discr-langevin}, makes a transition from configuration $y$
to configuration $x$ during the time step $\epsilon$; alternatively,
it can be seen as an approximation to the transition probability in
the continuous limit, correct to order $\O{\epsilon^2}$. In the
continuous limit the exact conditional probability is defined by the
relation: \begin{equation} P(x,\tau) = \int {\cal W}(x,y;\tau) P(y,0)
dy. \label{eq:cond-prob} \end{equation} By inserting this definition
into Eq. \eqref{fokker-planck}, one sees that ${\cal W}(x,y;\tau)$
satisfies Eq. \eqref{fokker-planck} with respect to $x$, subject to
the boundary condition: ${\cal W}(x,y;0)=\delta(x-y)$, {\it i.e.}
${\cal W}(x,y;\tau)$ is the Green's function of the Fokker-Planck
equation, Eq. \eqref{fokker-planck}. The Markovian character of random
walk, Eq. \eqref{discr-langevin}, allows one to define a simple
functional-integral representation for ${\cal W}(x,y;\tau)$ which
closely resembles the path-integral representation of the Green's
function of the time-dependent Schr\"odinger equation
\cite{feynman-hibbs}. As we will see in the following, this
resemblance is by no means accidental nor superficial.

Let $X_N=\{x_0,x_1,\cdots,x_N\}$ be a given random walk generated by
Eq. \eqref{discr-langevin}. Because of the Markovian character of the
chain, the corresponding probability density, $\PP_\epsilon[X_N]$
satisfies the relation: \begin{eqnarray} \PP_\epsilon[X_N] &\equiv&
{\tt Prob}[x(N\epsilon)=x_N; x((N-1)\epsilon)=x_{N-1}; \cdots ;
x(0)=x_0] \nonumber \\ &=& {\cal W}_\epsilon(x_N,x_{N-1}) \nonumber
{\tt Prob}[x((N-1)\epsilon)=x_{N-1}; \cdots ; x(0)=x_0] \nonumber \\
&=& \W _\epsilon(x_N,x_{N-1}) \PP_\epsilon[X_{N-1}], \end{eqnarray}
where $x(\tau)$ is the configuration of the system at time $\tau$. By
iterating this equation $N$ times, one obtains: \begin{equation}
\PP_\epsilon[X_N] = \W_\epsilon(x_N,x_{N-1}) \W_\epsilon( x_{N-1},
x_{N-2}) \cdots \W_\epsilon(x_1,x_0) P(x_0,0). \label{eq:Pe[X]}
\end{equation} The probability density that the system is in
configuration $x_N$ at time $\tau = N\epsilon$ is obtained from $
\PP_\epsilon[X_N]$ by integrating out the $N$ variables,
$\{x_0,\cdots,x_{N-1}\}$: \begin{eqnarray} P(x_N,N\epsilon) &=& \int
\PP_\epsilon[\{x_0,\cdots,x_N\}] dx_0 \cdots dx_{N-1}
\label{eq:w-integral_1} \\ &=& \int \W_\epsilon(x_N,x_{N-1}) \cdots
\W_\epsilon(x_1,x_0) P(x_0,0) dx_0 \cdots dx_{N-1}. \nonumber
\end{eqnarray} By comparing Eq. \eqref{cond-prob} with
Eq. \eqref{w-integral_1}, one obtains the desired functional-integral
representation for $\cal W$: \refstepcounter{equation}
\label{eq:w-integral_2} $$\displaylines{\quad \W(x,y;\tau) = \hfill
\cr \hfill \int \W_\epsilon(x,x_{N-1}) \W_\epsilon(x_{N-1},x_{N-2})
\cdots \W_\epsilon(x_1,y) ~dx_1 \cdots dx_{N-1},
\quad\eqref{w-integral_2}} $$ where $\epsilon= \tau/N$. The above
representation is {\it exact} for the {\it discrete} Markov chain
described by Eq. \eqref{discr-langevin}. In the continuous limit
($\epsilon \rightarrow 0$) it is understood that it holds by letting
$N=\tau / \epsilon \rightarrow \infty$, while keeping $\tau$ fixed.

\subsection{The classical-quantum mapping} \label{subs:cl-qu} In order
to establish the link between classical diffusion and imaginary-time
quantum evolution, we define a wave-function, $\Phi(x,\tau)$, through
the relation: \begin{equation} P(x,\tau) = \Phi_0(x) \Phi(x,\tau),
\label{eq:wvfct} \end{equation} where $\Phi_0(x) = \sqrt{P_s(x)}
\propto \ealla{-v(x)/2}$. By inserting Eq. \eqref{wvfct} into
Eq. \eqref{discr-FP} and dividing by $\Phi_0(x)$, we obtain:
\begin{equation} \Phi(x,\tau+\epsilon) = (1-\epsilon {\cal H})
~\Phi(x,\tau) + \O{\epsilon^2}, \label{eq:discr-Schr} \end{equation}
where: \begin{equation} {\cal H} = -{\partial^2 \over \partial x^2} +
{\cal V}(x), \label{eq:mapping-1} \end{equation} and \begin{eqnarray}
{\cal V}(x) &=& {1\over 4} \left ( {\partial v\over \partial x} \right
)^2 - \half ~{\partial^2 v\over \partial x^2} \nonumber \\ &=& {1\over
\Phi_0(x)} {\partial^2 \Phi_0(x) \over \partial x^2}~. 
\label{eq:mapping-2} \end{eqnarray} The continuous-time limit of
Eq. \eqref{discr-Schr} is formally equivalent to an imaginary-time
Schr\"odinger equation, \begin{equation} {\partial \Phi(x,\tau) \over
\partial \tau} = -{\cal H}~ \Phi(x,\tau), \label{eq:Schroedinger}
\end{equation} which could as well have been derived directly by
inserting Eq. \eqref{wvfct} into Eq. \eqref{fokker-planck}.

The wave-function $\Phi_0$ is a solution of Eq. \eqref{Schroedinger},
which means that it is an eigenfunction of the time-independent
Schr\"odinger equation, corresponding to a zero eigenvalue. A general
theorem of quantum mechanics states that the ground-state
eigenfunction of a Schr\"odinger equation with a local potential is
non-degenerate and node-less \cite{nodes}. Orthogonality with respect
to excited-state wave-functions implies that the ground state is the
only node-less eigenfunction. We thus arrive at the conclusion that
all the excited-state energies are strictly positive and that
$\Phi_0(x)$ is the only time-independent solution of Eq.
\eqref{Schroedinger} or, equivalently, that $P_s(x) \propto
\ealla{-v(x)}$ is the only stationary solution of the Fokker-Planck
equation \eqref{fokker-planck} \cite{ergodicity}. Furthermore, any
solution of Eq. \eqref{fokker-planck} would tend to $P_s$ for large
times, irrespective of the initial condition. If the spectrum of
${\cal H}$ has a gap, the approach to equilibrium is then
exponentially fast. This is easily seen by simple inspection. Consider
a system whose probability distribution at time $\tau=0$ is $P(x,0)$
and its expression in terms of the eigenfunctions and eigenvalues of
$\H$, which we indicate by $\Phi_n$ and $\E_n$: \begin{equation}
P(x,0) = \Phi_0(x)\sum_n c_n \Phi_n(x). \end{equation} The time
evolution of $P$ is readily derived from the time evolution of the
$\Phi$'s: \begin{equation} P(x,\tau)= c_0\Phi_0(x)^2 + \sum_{n\ne 0}
c_n\ealla{-\E_n\tau} \Phi_0(x) \Phi_n(x). \label{eq:P-time-evolution}
\end{equation} The normalization of $P(x,0)$ and the orthonormality of
the $\Phi$'s imply that $c_0=1$ and that the norm of $P(x,\tau)$ is
conserved. The above equation shows that the thermalization
time---{\it i.e.} the time necessary for the system to reach
equilibrium---is $\tau_0 \approx {1\over \E_1}$.

The fact that $P_s$ is the only stationary solution of the
Fokker-Planck equation, Eq. \eqref{fokker-planck}, implies that
classical expectation values over $P_s$---or, equivalently, quantum
expectation values over $\Phi_0(x)$---can be expressed as time
averages over the random walk, $x(\tau)$, generated by the Langevin
equation, Eq. \eqref{langevin}: \begin{equation} \int P_s(x) {\cal
A}(x)~dx \equiv \langle \Phi_0 | {\cal A} | \Phi_0 \rangle =
\lim_{T\to\infty} \lim_{t\to\infty} {1\over t} \int_T^{T+t} {\cal
A}\left ( x(\tau) \right ) ~d\tau. \label{eq:expct-value}
\end{equation}

A comparison between Eq. \eqref{master} and Eq. \eqref{discr-Schr}
allows one to establish a relation between the transition probability
of the classical random walk, ${\cal W}_\epsilon$, and the propagator
of the quantum system associated with it: \begin{equation} {\cal
W}_\epsilon(x,y) = \Phi_0(x) {\cal G}(x,y;\epsilon)/ \Phi_0(y)
+\O{\epsilon^2}, \label{eq:trans-prob} \end{equation} where:
\begin{equation} {\cal G} (x,y;\tau) \equiv \langle x |
\ealla{-\tau{\cal H}} | y \rangle. \label{eq:propagator}
\end{equation} The fact that the error in Eq. \eqref{trans-prob} is
indeed of second order in $\epsilon$, and not higher, can be proved by
pushing the Taylor expansion of Eq. \eqref{sega-2} to second order in
$\epsilon$ and by noting, for instance, that the second-order term
would give rise to a non-hermitian contribution to ${\cal G}$. Using
Eqs. \eqref{Pe[X]} and \eqref{trans-prob}, the probability density for
the random walk $X_N$ can be easily expressed in terms of a product of
quantum propagators, $\cal G$. Assuming that the system is at
equilibrium at $\tau=0$---{\it i.e.} that the probability distribution
for $x_0$ is $P_s(x_0)$---the probability distribution for the random
walk is: \begin{eqnarray} \PP_\epsilon[X_N] &=&
\underbrace{\Phi_0(x_N) {\cal G} (x_{N},x_{N-1};\epsilon) \cdots {\cal
G} (x_{1},x_{0};\epsilon) \Phi_0(x_0)}_{\displaystyle {\PP}[X_N]}
+\O{\epsilon} \nonumber \\ &\equiv& \PP[X] \times \QQ_\epsilon [X_N],
\label{eq:P[X]}
\end{eqnarray} where $\QQ_\epsilon [X_N] = 1 + \O{\epsilon}$ is a
time-discretization correction factor. The Markovian character of the
random walk implies a simple composition law for the $\PP$ probability
distributions: \refstepcounter{equation} \label{eq:composition}
$$\displaylines{\quad \PP[\{x_0,\cdots,x_k,\cdots,x_N\}] =
\PP[\{x_0,\cdots,x_k\}] \times \hfill \cr \hfill
\PP[\{x_k,\cdots,x_N\}] / P_s(x_k). \quad \eqref{composition}} $$
Inspection of Eq. \eqref{P[X]} shows that the probability distribution
${\PP}[X]$ is invariant under time reversal: \begin{eqnarray}
{\PP}[{\bar X}]={\PP}[X], \label{eq:time_reversal} \end{eqnarray}
where ${\bar X} \equiv \{x_N,\cdots,x_0\}$ is the path obtained from $X$
under time reversal.

\subsection{Classical vs. quantum time correlation functions}
Eq. \eqref{expct-value} expresses the identity between classical
thermal expectation values of local observables and quantum
expectation values of the same observables calculated over the ground
state of a suitably defined auxiliary system. Furthermore, these
expectation values can be expressed in terms of time averages over a
random walk. In the following we will see how this result can be
extended to time correlation functions. It will result that (real)
time correlation functions calculated over the random walk coincide
with the ground-state imaginary-time correlation functions of the
auxiliary quantum system.

The time auto-correlation function of an observable $\A(x)$ is
defined as: \begin{equation} \langle \A(\tau)\A(0) \rangle = \int \A(x)
\A(y) ~{\tt Prob}[x(0)=y; x(\tau)=x] ~dxdy. \label{eq:autocorr}
\end{equation} If the stochastic process is stationary ({\it i.e.} if
$P(x,\tau) = P_s(x)$), then one has:
\begin{eqnarray} {\tt Prob}[x(0)=y; x(\tau)=x] &=& \W(x,y;\tau) P_s(y)
\nonumber \\ &=&
\Phi_0(x)\G(x,y;\tau) \Phi_0(y). \end{eqnarray} By inserting this
relation into Eq. \eqref{autocorr}, the auto-correlation function
reads: \begin{eqnarray} \langle \A(\tau) \A(0) \rangle &=& \int
\A(x)\A(y) \Phi_0(x)\G(x,y;\tau) \Phi_0(y) dxdy \nonumber \\ &=&
\langle \Phi_0 | \ealla{\H\tau}\A\ealla{-\H\tau} \A | \Phi_0 \rangle
\nonumber \\ &\equiv& \langle \Phi_0 | \A(-i\tau) \A(0) | \Phi_0
\rangle , \label{eq:t-corre} \end{eqnarray} where we have used the fact that
$\ealla{\H\tau}|\Phi_0 \rangle = |\Phi_0\rangle$ and the definition
of a time-dependent operator in the Heisenberg representation: $\A(t)
= \ealla{i\H t} \A $ $\ealla{-i \H t}$.

Eq. \eqref{t-corre} can be used to express the autocorrelation time of
the observable $\A$ in terms of the spectral properties of the
auxiliary quantum Hamiltonian, $\H$. The autocorrelation time is
defined as the time integral of the normalized autocorrelation
function: \begin{equation} \tau_\A = \int_0^\infty \underbrace{\langle \A(\tau)
\A(0) \rangle - \langle \A \rangle^2 \over \langle \A^2 \rangle -
\langle \A \rangle^2}_{\displaystyle c_\A(\tau)}d\tau. \label{eq:tau_a}
\end{equation} The integrand in Eq. \eqref{tau_a} is defined in such
a way that $c_\A(0)=1$ and $c_\A(\infty)=0$. When $c_\A$ is
exponential, $\tau_\A$ is simply its decay constant: $c_\A(\tau) =
\ealla{-\tau/\tau_\A}$. In practical simulations, $\tau_\A$ determines
the statistical errors in the measure of $\A$:
\begin{equation} \langle \A \rangle \approx {1\over N}
\sum_{i=M+1}^{M+N} {\cal A}(x_i) \pm \sqrt{\Delta \A^2 {\tau_\A \over
N\epsilon}}, \label{eq:discr-expct-value} \end{equation} where $M$ is
chosen large enough so that equilibrium is reached ({\it i.e.}
$M>{1\over \epsilon E_1 }$), and $\Delta \A^2 \approx {1\over N-1}
\sum(\A(x_i) - \langle \A \rangle )^2$. In order to proceed further, we
first consider the spectral resolution of the quantum propagator,
$\G$: \begin{equation} \G(x,y;\tau) =\sum_n \ealla{-\E_n\tau}
\Phi_n(x) \Phi_n(y). \label{eq:spectral-resolution} \end{equation} By
inserting Eq. \eqref{spectral-resolution} into Eq. \eqref{t-corre},
the auto-correlation time can be easily cast into the form:
\begin{equation} \tau_\A = {1\over \Delta\A^2} \sum_{n\ne 0} { |
\langle \Phi_n | \A | \Phi_0 \rangle |^2 \over \E_n
}. \label{eq:t-corre-2} \end{equation} 

\section{From quantum mechanics back to classical diffusion}
The purpose of many quantum simulation techniques is to study the
ground-state properties of a system whose Hamiltonian is
\begin{equation} H = -{\partial^2 \over \partial x^2} + V(x),
\label{eq:hamiltonian} \end{equation} and whose (unknown) ground-state
wave-function and energy we indicate by $\Psi_0$ and $E_0$,
respectively. In the variational Monte Carlo method (VMC), an
approximate wave-function, $\Phi_0$, is used to generate a random walk
according to the discrete Langevin equation, Eq.
\eqref{discr-langevin}, with \begin{eqnarray} f(x) \equiv
f_{_{VMC}}(x) &=& -{\partial \over \partial x} \left ( -\log
\Phi_0(x)^2 \right ) \nonumber \\ &=& 2{1\over \Phi_0(x)} {\partial
\Phi_0(x) \over \partial x} , \label{eq:f_VMC} \end{eqnarray} and the
ground-state expectation value of the operator $\cal A$ is estimated
through Eq. \eqref{discr-expct-value}, where ${\cal A}(x) = {1\over
\Phi_0(x)}\A\Phi_0(x)$. Systematic errors in Eq.
\eqref{discr-expct-value} depend on the discretization of time and are
of order $\epsilon$. They can be eliminated in principle by enforcing
the detailed-balance condition on the master equation,
Eq. \eqref{master} \cite{smc}. A variational upper bound, ${\cal
E}_0$, to the ground-state energy can be estimated from
Eq. \eqref{discr-expct-value}: \begin{equation} {\cal E}_0 \equiv
\langle \Phi_0 | H | \Phi_0 \rangle \approx {1\over N} \sum_i {\cal
E}(x_i),
\label{eq:E_VMC} \end{equation} where \begin{equation} {\cal E}(x) =
{1\over \Phi_0(x)} H \Phi_0(x) \label{eq:local-energy} \end{equation}
is the so-called {\it local energy}. In the following we will show how the
{\it dynamical} properties of the random walk \eqref{langevin} can be
used to systematically improve upon this VMC procedure and to estimate,
{\it exactly} within statistical noise, the ground-state properties of
quantum systems.

Let us first observe that the trial wave-function, $\Phi_0$,
implicitly defines a reference (unperturbed) Hamiltonian, $H_0$, whose
{\it exact} ground state is $\Phi_0$. The potential function which
defines $H_0$ is simply obtained by inverting the time-independent
Schr\"odinger equation: \begin{equation} V_0(x) = {1\over \Phi_0(x)}
{\partial^2 \Phi_0(x) \over \partial x^2} + \E_0. \label{eq:v0}
\end{equation} A comparison between Eq. (\ref{eq:v0}) and
Eqs. (\ref{eq:mapping-1},\ref{eq:mapping-2}) shows that the potential
obtained from the inversion of the Schr\"odinger equation coincides
with the effective quantum potential resulting from the
classical-quantum mapping discussed in Sec. \ref{subs:cl-qu}, provided
that the classical random walk, Eqs.
(\ref{eq:langevin},\ref{eq:discr-langevin}) is driven by the VMC force
defined in Eq. \eqref{f_VMC}. The original Hamiltonian, Eq.
\eqref{hamiltonian}, can then be cast into the form: \begin{equation}
H = \underbrace{-{\partial^2 \over \partial x^2} + V_0(x)
}_{\displaystyle \cal H} + \underbrace{\E(x) -\E_0 }_{ \displaystyle
\Delta\H}. \end{equation} By construction, $\Phi_0$ is the
ground-state of $\cal H$, and ${\cal E}(x) = E_0$ if $\Phi_0$ is the
ground state of $H$. When $\Phi_0$ is not the ground state of $H$, the
local energy ${\cal E}(x)$ is not a constant, and the ground-state
properties of $H$ can in principle be obtained by perturbation theory
with respect to $\Delta\H$.

Let us calculate the corrections to the ground-state energy, $\E_0$,
up to second order in $\Delta\H$. The first-order correction vanishes
by construction, given that $\langle\Phi_0 | \E | \Phi_0 \rangle =
\E_0$. The second-order correction is given by:
\begin{equation} \Delta\E_0^{(2)} = - \sum_{n\ne 0} {| \langle \Phi_0
| \E | \Phi_n \rangle |^2 \over \E_n}. \label{eq:deltaE2}
\end{equation} By comparing this expression with
Eq. \eqref{t-corre-2}, the second-order correction to the ground-state
energy can be expressed in terms of the fluctuations of the local
energy and of its auto-correlation time: \begin{equation}
\Delta\E_0^{(2)} = \tau_\E \Delta\E^2. \end{equation} In the
following, we show how this relation between perturbative corrections
to the ground-state energy and (integrals of) auto-correlation
functions of the local energy along the random walk can be generalized
to arbitrary order. The resulting perturbation series can be
effectively summed to infinite order by an appropriate re-sampling of
the random walks generated by the VMC Langevin equation, Eqs.
(\ref{eq:langevin},\ref{eq:discr-langevin},\ref{eq:f_VMC}).

\subsection{Stochastic perturbation theory}

Given the Hamiltonian $H$, Eq. (\ref{eq:hamiltonian}), and a trial
wave-function, $\Phi_0$, which we suppose to be non-orthogonal to its
ground state, $\Psi_0$, the ground-state energy of H can be expressed
as: \begin{eqnarray} E_0 = \lim_{\tau\to\infty} {\langle \Phi_0 | H
{\rm e}^{-H\tau} | \Phi_0 \rangle \over \langle \Phi_0 | {\rm
e}^{-H\tau} | \Phi_0 \rangle} &=& -\lim_{\tau\to\infty} {d\over
d\tau}\log\langle \Phi_0 | {\rm e}^{-H\tau} | \Phi_0 \rangle
\label{eq:ground-state-1} \\ &=& -\lim_{\tau\to\infty} {1\over
\tau}\log\langle \Phi_0 | {\rm e}^{-H\tau} | \Phi_0 \rangle. 
\label{eq:ground-state-2} \end{eqnarray} These equations are
easily demonstrated by expressing $\Phi_0$ as a linear combination of
the eigenfunctions of $H$, $\{\Psi_n\}$: $\Phi_0 = \sum_n c_n \Psi_n$,
and by inserting this expression into
Eqs. (\ref{eq:ground-state-1},\ref{eq:ground-state-2}):
\begin{equation} \log \langle \Phi_0 | {\rm e}^{-H\tau} | \Phi_0
\rangle = -E_0\tau + \log \left (c_0^2 \right ) + \O{\ealla{-(E_1-E_0)
\tau}}. \label{eq:logeH} \end{equation} The following basic identity
holds: \begin{eqnarray} \Z_0 \equiv \langle \Phi_0 | {\rm e}^{-H\tau}
| \Phi_0 \rangle &=& \int \ealla{-\SS [X]} \PP [X] ~\DD [X] \nonumber
\\ &\equiv & \left \langle \ealla{-\SS [X]} \right \rangle,
\label{eq:feynman-kac} \end{eqnarray} where $\PP [X]$ is the
probability distribution for the random walk $X$, Eq. \eqref{P[X]},
$\DD [X] = dx_0 \cdots dx_N$, and the action $\SS [X]$ is a
functional of the random walk which in the continuum limit coincides
with the time integral of the local energy: \begin{eqnarray} \SS
[X] &=& \epsilon\sum_{i=1}^N{\cal E}(x_i) + \O{\epsilon} \nonumber
\\ &\approx& \int_0^\tau {\cal E}(x(\tau')) d\tau'. \label{eq:action}
\end{eqnarray} Eq. \eqref{feynman-kac} is a generalization of the
Feynman-Kac formula \cite{feynman-kac} and can be demonstrated as
follows. \begin{equation} \Z_0 = \int \Phi_0(x) G(x,y;\tau)
\Phi_0(y)~dxdy, \label{eq:vac-1} \end{equation} where
\begin{equation} G(x,y;\tau) = \langle x | {\rm e}^{-H\tau} | y
\rangle \label{eq:propagator-1}\end{equation} is the imaginary-time
propagator of the full Hamiltonian. We now break the propagator in
Eq. (\ref{eq:vac-1}) into the product of $N$ short-time propagators:
\begin{equation} \Z_0= \int
\underbrace{\Phi_0(x_0) G(x_0,x_1;\epsilon) \cdots G(x_{N-1},x_N;\epsilon)
\Phi_0(x_N)}_{\displaystyle {\bf P}[X]} ~\DD [X]. 
\label{eq:vac-2} \end{equation}
Eqs. \eqref{feynman-kac} and \eqref{vac-2} can be seen as a
{\it definition} of the action: \begin{equation} {\bf P}[X] = \PP [X]
\ealla{-\SS [X]}. \label{eq:P[X]-1} \end{equation} By using the
Trotter formula, \begin{equation} G(x,y;\epsilon) = {\cal G}(x,y;\epsilon)
\ealla{-\epsilon~ {\cal E}(y)} + \O{\epsilon^2}, \label{eq:trotter}
\end{equation} one sees that the action defined by Eq. \eqref{P[X]-1} can
indeed be expressed by the time integral given by Eq. \eqref{action}. 

$\Z_0$ plays the role of a pseudo partition function, in the sense
that the energy of the system---as well as other observables, as we
will see---can be calculated by differentiating it much in the same
way as one would do in classical statistical mechanics. In particular,
the two expression \eqref{ground-state-1} and \eqref{ground-state-2}
for the ground-state energy correspond to the zero-temperature limits
of the internal and free energies, respectively. By inserting
Eq. \eqref{feynman-kac} into Eq. \eqref{ground-state-2}, one could
derive a systematic perturbative expansion of the ground-state energy
in powers of $\E$. Each term of the series is basically a cumulant of
the action which in turn can be seen as the integral of a suitably
defined {\it connected} time correlation function of the local energy,
calculated along the Langevin random walk. This kind of {\it
stochastic} perturbation theory can be effectively carried on to
infinite order by inserting Eq. \eqref{feynman-kac} into
Eq. \eqref{ground-state-1}. The final result reads: \begin{eqnarray}
E_0 &=& \lim_{\tau\to\infty} { \left \langle {\cal E}(x(\tau))
\ealla{-\SS [X]} \right \rangle \over \left \langle \ealla{-\SS 
[X]} \right \rangle} \nonumber \\ &\equiv& \lim_{\tau\to\infty}
\llangle {\cal E}(x(\tau)) ~\rrangle, \label{eq:ground-state-3}
\end{eqnarray} where the double bracket $\llangle \cdot \rrangle$
indicates that the average over the random walks ({\it quantum paths})
is re-weighted by the exponential of the action, Eq. \eqref{action}.

Eq. \eqref{ground-state-3} can be turned into an algorithm for
calculating the ground-state energy. The calculation would proceed as
in a standard VMC simulation, with the difference that the local
energy must be weighted with the exponential of the action,
$\ealla{-\SS}$, calculated along a segment of the random walk long
$\tau$ and ending at the time when the measure is taken. This
algorithm, which was first proposed in Ref. \cite{caffarel}, is bound
to fail in all those case where the number of particles is so large or
the quality of the trial wave-function is so poor that the
fluctuations of the action make the weighting procedure
impractical. As a matter of fact, the {\it pure-diffusion Monte Carlo}
(PDMC) of Ref. \cite{caffarel} has never been applied but to very
simple quantum systems. In the next session we will show how the
fluctuations of the action can be effectively dealt with through a new
algorithm based on importance sampling. Before doing this, we want to
show how the ideas exposed so far can be exploited to calculate
general ground-state expectation values and response functions.

\subsection{Calculation of observables} Other observables can be
calculated along similar lines starting from the Hellman-Feynman
theorem \cite{hellman-feynman}: \begin{equation} \langle\Psi_0 | \A |
\Psi_0 \rangle = \left . {d E_\lambda \over d\lambda} \right
|_{\lambda=0},\end{equation} where $E_\lambda$ is the ground-state
energy of a system whose Hamiltonian is \begin{equation} H_\lambda =
H+\lambda \A. \label{eq:h_lambda} \end{equation} By using 
Eq. \eqref{ground-state-2}, $E_\lambda$ can be put in the form:
\begin{equation} E_\lambda = -\lim_{\tau\to \infty}{1\over\tau} \log\left
\langle \ealla{-\SS_\lambda[X]} \right \rangle, \label{eq:e_of_lambda}
\end{equation} where \begin{equation} \SS_\lambda[X] = \int_0^\tau
({\cal E}(x(\tau')) + \lambda {\cal A}(x(\tau')))~d\tau'. 
\end{equation} The expectation value of $\A$ can then be expressed as:
\begin{eqnarray} \langle \Psi_0|\A| \Psi_0 \rangle &=&
\lim_{\tau\to\infty} { \left \langle {1\over \tau}\int_0^\tau {\cal
A}(x(\tau')) d\tau' {\rm e}^{-\int_0^\tau {\cal E}(\tau')d\tau'}
\right \rangle \over \left \langle {\rm e}^{-\int_0^\tau {\cal
E}(\tau')d\tau'} \right \rangle} \nonumber \\ &\equiv&
\lim_{\tau\to\infty} \llangle {1\over\tau} \int_0^\tau {\cal
A}(x(\tau'))~d\tau' \rrangle \label{eq:vave}
\end{eqnarray}

\subsection{Response functions}

A simple extension of the ideas used in the previous section leads to
a technique for evaluating response functions. Let us suppose that the
system is coupled to a set of external variables, $\{\lambda_i\}$
through the operators $\{\A_i\}$: \begin{equation} H_{\{\lambda\}} = H
+ \sum_i \lambda_i \A_i. \end{equation} In the case of an external
potential, for instance, the index $i$ labels the coordinate, $x$,
$\lambda_i$ is the potential itself, $V_{ext}(x)$, and $\A_i$ is the
particle-density operator, $n(x)$. We define a generalized
susceptibility, $\chi$, as the derivative of the expectation value of
one of the $\A$ operators with respect to one of the $\lambda$'s:
\begin{eqnarray} \chi_{ij} &=& {\partial\langle \A_i \rangle \over
\partial \lambda_j } \nonumber \\ &=& {\partial^2 E_\lambda \over
\partial \lambda_i \partial \lambda_j}. \end{eqnarray} By using
Eq. \eqref{e_of_lambda}, $\chi$ can be put in the form:
\begin{eqnarray} \chi_{ij} &=& -\lim_{\tau\to\infty} {1\over \tau}
\left [ \llangle \left ( \int_0^\tau \A_i(\tau') d\tau' \int_0^\tau
\A_j(\tau') d\tau' \right ) \rrangle \right . \nonumber \\ &&
\quad\quad\quad\quad \quad\quad
\left . - \llangle \int_0^\tau \A(\tau')
d\tau' \rrangle^2 \right ] \nonumber \\ &=& -\lim_{\tau\to\infty}
{1\over \tau} \llangle \int_0^\tau d\tau_1 \int_0^\tau d\tau_2
(\A_i(\tau_1) \A_j(\tau_2) -\bar{\A_i} ~\bar{\A_j} )
\rrangle, \label{eq:corr} \end{eqnarray} where $\bar{\A}$ is the
time average of $\A$ over the random walk: $\bar{\A} = 
{1\over \tau} \int_0^\tau \A(\tau') d\tau'$. We now split the domain
of integration $[0\leq \tau_1 \leq \tau; 0\leq \tau_2 \leq \tau ]$
into two sub-domains $[\tau_1<\tau_2]$ and $[\tau_2 < \tau_1]$, and
change the variables of integration $ \{\tau_1,\tau_2\} \rightarrow
\{\tau_1,\tau_2-\tau_1\}$ and $ \{\tau_1,\tau_2\} \rightarrow
\{\tau_1-\tau_2,\tau_2\}$ in the two sub-domains respectively. In the
large $\tau$ limit, the susceptibility can then be cast into the form:
\begin{equation} \chi_{ ij} = - \llangle \int_0^\tau [c_{ij}(\tau') +
c_{ji}(\tau') ] d\tau' \rrangle, \end{equation} where the time
auto-correlation function is defined by: \begin{equation} c_{ij}(\tau)
={1\over \tau} \int_0^\tau \A_i(\tau') \A_j(\tau') d\tau' -{1\over
\tau^2} \int_0^\tau \A_i(\tau') d\tau' \int_0^\tau \A_j(\tau')
d\tau'. \end{equation} The symmetrized time correlation function
$c_{ij} + c_{ji}$ is a functional of the path, whose time integral
({\it i.e.} whose $\omega=0$ Fourier component) provides an estimator
for the static response function. This argument can be easily
generalized to show that other ($\omega \ne 0$) Fourier components of
the same time correlation function provide suitable estimators for the
dynamic susceptibility.

\section{The algorithm}

In the previous section we have shown that the calculation of ground-state
expectation values and response functions can be reduced to the weighted
average of suitable estimators over the space of random walks. A
straightforward evaluation of these averages over a Langevin trajectory
would be very inefficient because the weight, $\ealla{-\SS}$, may
vary a lot. This problem can be solved by converting the {\it weighting}
of the quantum paths into a {\it re-sampling} of them, through an
appropriate Metropolis algorithm \cite{metropolis}. As a by-product, the
use of the Metropolis algorithm allows one to reduce the systematic errors
due to the discretization of time in an efficient and convenient way.

Let $G^{(n)}$ be any approximation to the full imaginary-time
propagator, Eq. \eqref{propagator-1}, correct to order $n$ in
$\epsilon$, and ${\bf P}^{(n)}[X]$ the corresponding approximation to
${\bf P}[X]$, Eq. \eqref{vac-2}, which will be affected by errors of
order $n$. Let us also define the corresponding approximations for the
unperturbed propagator, ${\cal G}^{(n)}$, Eq. \eqref{propagator}, path
probability distribution, $\PP^{(n)}$, and time-discretization
correction factor, $\QQ_\epsilon^{(n)}$, Eq. \eqref{P[X]}, and action,
$\SS^{(n)}[X]$, Eq. \eqref{P[X]-1}. For instance, one can take:
\begin{eqnarray} G^{(2)}(x,y;\epsilon) &=& {1\over \sqrt{4\pi
\epsilon}} \ealla{-{(x-y)^2 \over 4\epsilon} - {\epsilon \over 2}
(V(x) + V(y) )} = G(x,y;\epsilon) + \O{\epsilon^3}, \nonumber \\ {\cal
G}^{(2)}(x,y;\epsilon) &=& {1\over \sqrt{4\pi \epsilon}}
\ealla{-{(x-y)^2 \over 4\epsilon} - {\epsilon \over 2} \left
({\Phi_0''(x)\over \Phi_0(x)} + {\Phi_0''(y)\over \Phi_0(y)} \right )
} = {\cal G}(x,y;\epsilon) + \O{\epsilon^3}, \nonumber \\ \PP^{(2)}[X]
&=& \Phi_0(x_N) {\cal G}^{(2)}(x_N,x_{N-1};\epsilon) \cdots {\cal
G}^{(2)}(x_1,x_0;\epsilon) \nonumber \Phi_0(x_0) \\ &=& \PP [X]
+\O{\epsilon^2},\\ \SS^{(2)}[X] &=& \ealla{-{\epsilon \over 2}
\sum_{i=1}^N \left ( {\cal E}(x_i) + {\cal E}(x_{i-1}) \right )} =
\SS[X] + \O{\epsilon^2} . \nonumber\end{eqnarray} Any ground-state
expectation value or response function can be put in the form:
\begin{equation} \llangle {\bf F} [X] \rrangle = {\int {\bf P}^{(n)}[X]
~{\bf F}[X] ~\DD[X] \over \int {\bf P}^{(n)}[X] ~\DD[X] } +
\O{\epsilon^n}. \end{equation}

In order to calculate the above expectation value through the
Metropolis method, it is necessary to construct a Markov chain of
random walks so that the corresponding transition probability, ${\bf
W}[Y\leftarrow X]$, satisfies detailed balance: \begin{equation} {\bf
W}[Y\leftarrow X] \times {\bf P}^{(n)}[X] = {\bf W}[X\leftarrow Y]
\times {\bf P}^{(n)}[Y] \label{eq:detailed-balance} \end{equation} In
the Metropolis algorithm, the transition probability ${\bf W}$ is
split into the product of an {\it a-priori} transition probability,
${\bf W}^0$, times a probability, ${\bf A}$, that a move {\it
proposed} according to ${\bf W}^0$ is accepted: ${\bf W}[Y\leftarrow
X] = {\bf W}^0[Y\leftarrow X] \times {\bf A}[Y\leftarrow X]
$. Detailed balance, Eq. \eqref{detailed-balance}, can be satisfied by
choosing \cite{gen-metropolis}: \begin{equation} {\bf A}[Y\leftarrow
X] = {\rm min}(1, {\bf R}[Y\leftarrow X]),
\end{equation} where \begin{equation} {\bf R}[Y\leftarrow X] = { {\bf
W}^0[X\leftarrow Y] \times {\bf P}^{(n)}[Y] \over {\bf
W}^0[Y\leftarrow X] \times {\bf P}^{(n)}[X] }. \label{eq:R}
\end{equation} Given a quantum path, $X\equiv \{x_0,\cdots,x_N\}$, we
propose a new path, $Y$, by propagating the random walk forward by
$M<N$ steps, according to the VMC Langevin equation,
Eqs. (\ref{eq:discr-langevin},\ref{eq:f_VMC}): $Y\equiv
\{x_M,\cdots,x_N,\cdots,x_{M+N}\}$. The corresponding a-priori
transition probability is: \begin{eqnarray} {\bf W}^0[Y\leftarrow X]
&=& \W_\epsilon(x_{N+M},x_{N+M-1}) \cdots \W_\epsilon(x_{N+1},x_{N})
\\ &=& \PP^{(n)}[\{x_{N},\cdots, x_{N+M}\}]
\QQ^{(n)}_\epsilon[\{x_{N},\cdots, x_{N+M}\}] / P_s(x_N) , \nonumber
\end{eqnarray} where we have used Eq. \eqref{P[X]}. By inserting this
expression for the a-priori probability into Eq. \eqref{R}, the latter
can be cast into the form: \refstepcounter{equation}
$$\displaylines{\quad {\bf R}[Y\leftarrow X] = {
\PP^{(n)}[\{x_M,\cdots,x_{0}\}] \times
\QQ^{(n)}_\epsilon[\{x_M,\cdots,x_{0}\}] / P_s(x_M) \over
\PP^{(n)}[\{x_N,\cdots,x_{N+M}\}] \times
\QQ^{(n)}_\epsilon[\{x_N,\cdots,x_{N+M}\}] / P_s(x_N) } \hfill \cr
\hfill \times \PP^{(n)}[\{x_0,\cdots,x_{N}\}] \times
{\ealla{-\SS^{(n)}[X] }\over \PP^{(n)}[\{x_M,\cdots,x_{N+M}\}] \times
\ealla{-\SS^{(n)}[Y] }}. \quad (\theequation)}$$ With the aid of the
composition law for the random-walk probability distributions,
Eq. \eqref{composition}, and of the time--reversal property, Eq. 
\eqref{time_reversal}, the above equation finally reads:
\begin{equation} {\bf R}[Y\leftarrow X] = \ealla{-(\SS^{(n)}[Y] -
\SS^{(n)}[X])} \times { \QQ^{(n)}_\epsilon[\{x_M,\cdots,x_{0}\}]
\over {\QQ}^{(n)}_\epsilon[\{x_N,\cdots,x_{N+M}\}] }. 
\label{eq:R[YX]}
\end{equation}
Notice that the ratio of the $\QQ$'s goes to one in the continuous-time
limit ($\epsilon\rightarrow 0$) and that for finite $\epsilon$ it can
be explicitly calculated, providing thus a systematic way for reducing
to any desired order the time-discretization errors.

The dynamical variables of our simulations are quantum paths (or
segments of random walks) which can be formally associated with
polymers, much in the same spirit as this is done in path-integral
simulations. The polymer dynamics which would correspond to our
algorithm is known in the literature as {\it reptation}
\cite{reptation}. For this reason, we name our algorithm {\it
Reptation Quantum Monte Carlo} (RQMC).

The practical implementation of RQMC is extremely simple,
at the level of a VMC simulation. The algorithm can 
be summarized as follows:

\begin{enumerate}

\item Using Eqs. \eqref{discr-langevin} and \eqref{f_VMC}, generate a
random walk long enough that its end point, $x(T)$, is distributed
according to $\Phi_0^2$.

\item Generate a further segment of random walk corresponding to the
time interval [$T,T+\tau$]. $\tau=N\epsilon$ should be large enough
that the limit $\tau\to\infty$ in Eqs. \eqref{ground-state-3},
\eqref{vave}, and \eqref{corr} is reached to the desired accuracy. Set
$X\equiv \{x_0,x_1,\cdots,x_N\} \leftarrow \{ x(T), x(T+\epsilon),
\cdots,x(T+\tau)\}$.

\item Select a `direction of time' ({\it forward} or {\it backward})
with equal probability. If the choice is {\it backward}, set $X
\leftarrow \bar X = \{x_N,x_{N-1},\cdots,x_0\}$. This step is
necessary to ensure the micro-reversibility of the algorithm.

\item Generate a segment of random walk corresponding to the time
interval $[T+\tau,T+\tau+\delta]$ and set $Y\equiv
\{x_M,x_{M+1},\cdots,x_{N+M}\} \leftarrow \{ x(\delta),
x(T+\delta+\epsilon), \cdots,x(T+\tau+\delta)\}$. The value of
$\delta$ is sampled from an uniform deviate in the interval
$[0,\Delta]$ whose width is chosen so as to minimize the
auto-correlation times of the measured quantities. Sampling $\delta$
instead of keeping it constant helps to avoid that the path remains
occasionally stuck at a fixed position for a long time.

\item Evaluate ${\bf R}[Y \leftarrow X ]$ according to
Eq. \eqref{R[YX]}.

\item If ${\bf R} > 1$, set $X \leftarrow Y$. If ${\bf R} < 1$, set $X
\leftarrow Y$ with probability $\bf R$.
 
\item Accumulate the ground-state energy and other observables using
appropriate estimators (see the next section for an optimal choice of
the estimators).

\item Go to 3.

\end{enumerate}

This algorithm has been preliminary tested for the hydrogen atom
using an approximate trial function. Exact results for the average of
several moments of electron--nucleus distance have been reproduced
within a statistical error pushed down to a small fraction of the
difference between the exact value and the extrapolated estimate
(i.e. twice the mixed average minus the variational average
\cite{ceperley_79}).

\section{A case study of $^4$He} We now discuss the calculation of
several properties of superfluid $^4$He, with the purpose of showing
that the method can be successfully applied to systems of actual
physical interest. Based on the limited experience gained in this case
study, we also present some performance comparisons with related
techniques.

\subsection{Details of the simulation}
We consider $N_P=64$ $^4$He atoms interacting through a pair
potential, as obtained from first-principles calculations
\cite{korona}. The simulation was performed in a cubic box with
periodic boundary conditions at the experimental equilibrium density,
$\rho = 0.02186$ \AA$^{-3}$. 
The trial function $\Phi_0$ includes pair and nearly optimal
three--body correlations \cite{euler}. This is a relatively good trial
function. The variational energy is less than $0.3\rm ~K$ above the exact
ground state energy, whereas the variational bias is larger than $1.1\rm ~K$
using pair correlations only. 
The quantities we compute are total and
potential energies, the imaginary-time correlations 
of the density fluctuation operators,
$\rho_q=\sum_i\exp(-i{\bf q}\cdot{\bf r}_i)$:
\begin{equation}
F(q,\tau) = \langle \rho_q(\tau) \rho_{-q}^\dagger(0)\rangle/N_{P},
\end{equation} and the diffusion
coefficient of the center of mass motion,
\begin{equation} D(\tau)=\langle[{\bf r}_{CM}(\tau)-{\bf
r}_{CM}(0)]^2\rangle N_P/(6\tau). \label{eq:cm} \end{equation} 

The parameters of our simulations are as follows. The time step is
$\epsilon=0.001\rm ~K^{-1}$, which gives a systematic bias of the
order of $10^{-2}\rm ~K$ on the total energy.  For the calculation of
total and potential energy the length of the path was $\tau=0.4\rm
~K^{-1}$, corresponding to $N=400$ time slices. The calculation of
imaginary-time correlations over a significant range required the use
of longer paths, up to $N=700$. The value of the energy resulting from
the simulation with such longer paths confirmed that the finite-$\tau$
bias in the results obtained with $N=400$ was smaller than statistical
errors. Note that the length of the path adversely affects the
efficiency, because the relaxation time of the polymer in the
reptation algorithm is proportional to $N^2$ \cite{lebowitz}.  The
number of time slices of each reptation move is uniformly sampled
between 0 and 20, yielding an acceptance ratio of $\approx 80\%$.

\begin{figure} \epsfxsize=12cm \begin{center} \leavevmode
\epsffile{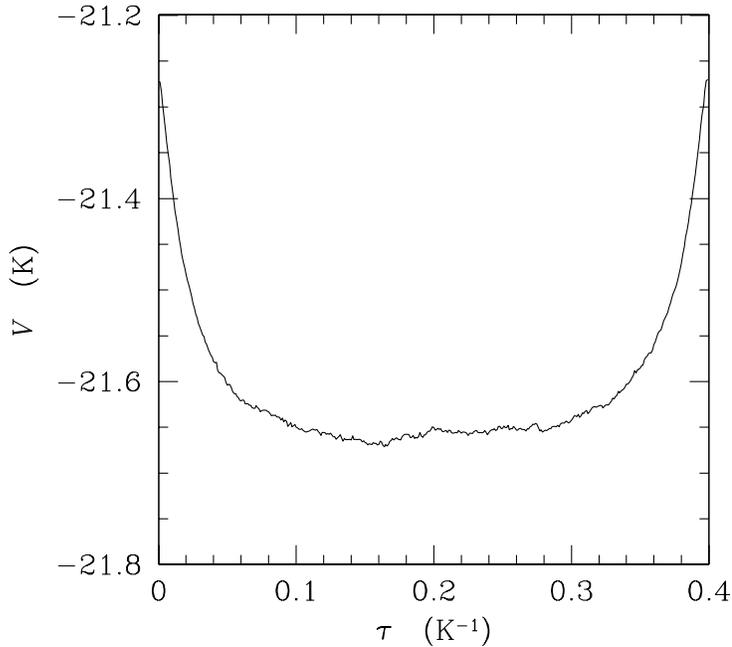} \end{center} \caption[]{Average of the potential
energy on individual time slices along the path. The statistical
error on the central slices is $\simeq 0.03\rm ~K$. This result was
obtained using a trial function with pair correlations only: note that
$V$ converges to the same value given in Table 1, obtained using a
trial function with pair and triplet correlations.} \label{fig:v}
\end{figure}

Fluctuations in the average of the total energy are reduced using a
symmetrized form of Eq. \eqref{ground-state-3}, {\it i.e.} accumulating
$\left[{\cal E}(x(\tau))+{\cal E}(x(0))\right]/2$.  Expectation values
of local observables are computed averaging time integrals along the
path, Eq. \eqref{vave} and Eq. \eqref{corr}.  Although these
expressions are exact in the $\tau\to\infty$ limit, the contributions
coming from the extrema of the path are clearly biased.  For instance,
the average of $\A$ in the initial or final time slices ($\left\langle
\A(x(0)) \right\rangle$ and $\left\langle \A(x(\tau)) \right\rangle$)
yields the mixed estimate: \begin{eqnarray} \lim_{\tau\to\infty} {
\left \langle \A (x(0)) {\rm e}^{-\int_0^\tau {\cal E}(\tau')d\tau'}
\right \rangle \over \left \langle {\rm e}^{-\int_0^\tau {\cal
E}(\tau')d\tau'} \right \rangle} = \lim_{\tau\to\infty} { \langle
\Phi_0 | \A {\rm e}^{-H\tau} | \Phi_0 \rangle \over \left \langle
\Phi_0 | {\rm e}^{-H\tau} | \Phi_0 \right \rangle} = {\langle
\Phi_0|\A| \Psi_0 \rangle \over \langle \Phi_0|\A| \Psi_0 \rangle}.
\label{eq:mixed} \end{eqnarray} On the other hand any time slice a
distance ${\bar \tau}$ apart from the extrema, such that $\exp(-H{\bar
\tau})|\Phi_0\rangle \simeq |\Psi_0\rangle$, gives an unbiased
contribution to the time integral of $\A$.  Therefore it is convenient
to restrict the time integral of $\A$ in Eqs.  \eqref{vave} and
\eqref{corr}, to the inner section of the path, where the bias is
reduced.  For the potential energy and the imaginary-time correlations
we exclude the contributions from 150 time slices on each side of the
path. This is a rough estimate of the time it takes for the average
potential energy to converge within a few hundredths K from its value
at slice 0 or $N_\tau$ (which is by Eq. \eqref{mixed} the mixed
estimate) to the unbiased estimate. This is demonstrated by the
average of the potential energy on individual time slices, shown in
Fig. \ref{fig:v}. We have not studied the corresponding convergence
times for the density--density correlations.

\begin{table} \begin{center} \caption[]{Ground state energy, $E_0$, and
potential energy, $V$, as computed from RQMC and traditional diffusion
Monte Carlo runs of 3$\times$10$^6$ Monte Carlo steps. Units are
K. The length of the path in the RQMC calculation is $\tau=0.4\rm
~K^{-1}$, and the length of the forward walk for $V$ in the diffusion
Monte Carlo calculation is $0.2\rm ~K^{-1}$.}
\begin{tabular}{ccc}
           &    $E_0$    &    $V$      \\
\hline
 RQMC      & -7.4066(27) & -21.644(15) \\
 BDMC       & -7.3902(15) & -21.674(21) \\
\hline
\end{tabular}
\label{tab:table1} 
\end{center}
\end{table}

\subsection{Results} Our results for the total and potential energies
are listed in Table 1.  The inter-particle potential adopted 
\cite{korona} overestimates the experimental binding energy of 
$-7.17 \rm ~K$ because of the neglect of three--body forces 
(mostly triple--dipole repulsion).
Also reported in Table 1 are the corresponding data obtained
from a standard Branching Diffusion Monte Carlo (BDMC) calculation using 
the same time step and trial function. The small differences between 
the results of the two algorithms have to be attributed to different 
time step biases.

\begin{figure} \epsfxsize=12cm \begin{center} \leavevmode
\epsffile{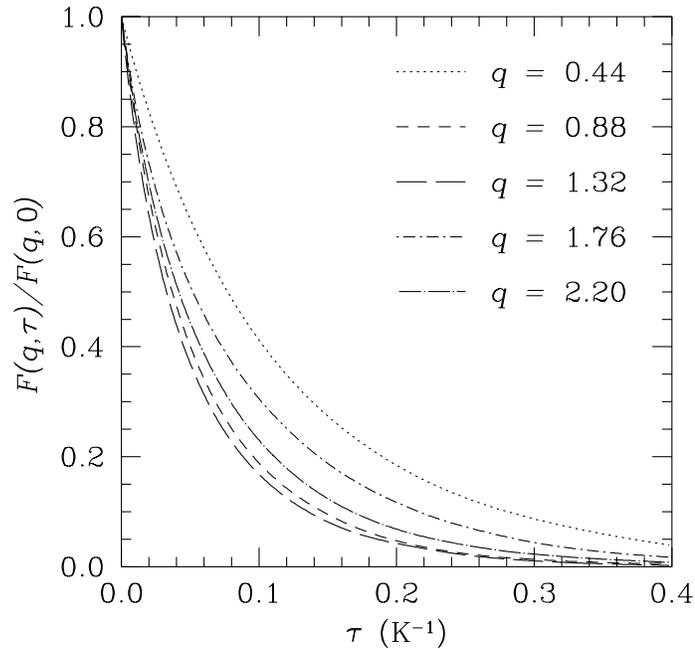} \end{center} \caption[]{Imaginary-time correlations
of the density fluctuation operator. Averages are taken on the inner
part of a path of length $0.7\rm ~K^{-1}$. The statistical error
ranges from approximately 0.5\% at $\tau=0$ up to 5\% at
$\tau=0.4\rm ~K^{-1}$.}
\label{fig:corf} \end{figure}

In figure \ref{fig:corf} we show the density-density correlation
function, $F(q,\tau)$, for a few values of $q$, as obtained from a run
of $10^7$ MC steps, which required about one week CPU time on a
workstation.  Note that at virtually no additional
cost many more $q$ vectors, belonging to the reciprocal lattice of the
simulation box, could have been included in the calculation.

\begin{figure} \epsfxsize=12cm \begin{center} \leavevmode
\epsffile{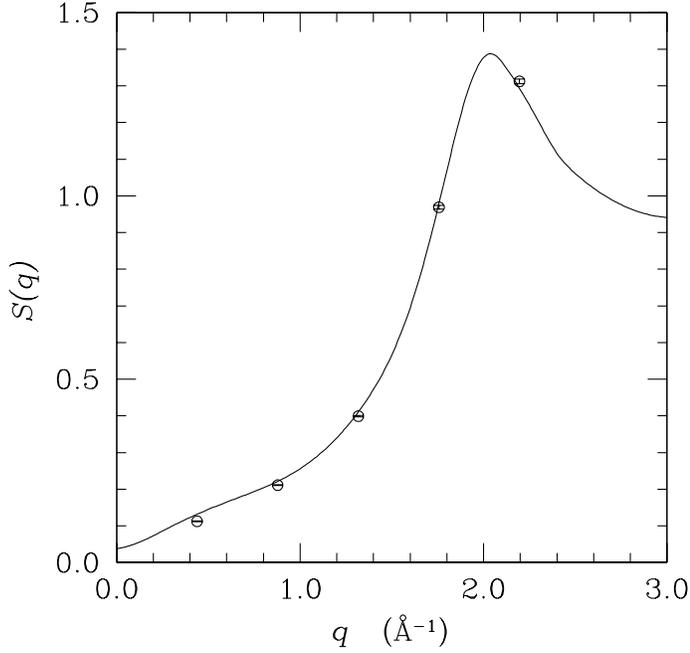} \end{center} \caption[]{Static 
structure factor at the five wave vectors listed in Fig. 
\ref{fig:corf} (open circles). The solid line is the experimental 
$S(q)$ measured by neutron scattering \cite{sears}.} \label{fig:sq} 
\end{figure}

$F(q,\tau)$ is related to several quantities of physical interest
\cite{pines}, including the static structure factor,
\begin{eqnarray} S(q)=F(q,0)=\int_0^\infty d\omega~ S(q,\omega),
\end{eqnarray} the static linear response function,
\begin{eqnarray} \chi(q)=-2\int_0^\infty d\tau~
F(q,\tau)=-2\int_0^\infty d\omega~ S(q,\omega)/\omega, \end{eqnarray}
and the dynamical structure factor, \begin{eqnarray}
F(q,\tau)=\int_0^\infty d\omega ~\ealla{-\omega\tau} ~S(q,\omega).
\label{eq:sqom} \end{eqnarray}

\begin{figure} \epsfxsize=12cm \begin{center} \leavevmode
\epsffile{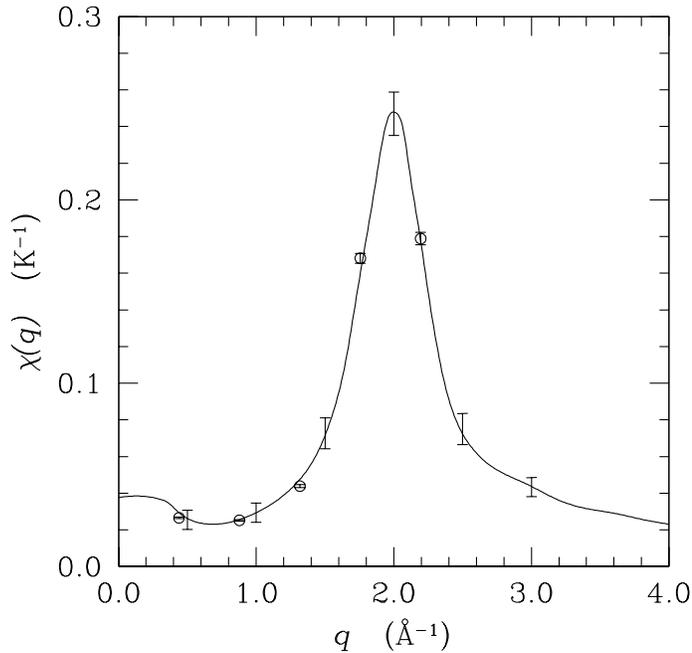} \end{center} \caption[]{The static response function
at the five wave vectors listed in Fig. \ref{fig:corf} (open
circles). The solid line is the experimental result of
Ref. \cite{woods}.} \label{fig:chiq} \end{figure}

Low moments of $S(q,\omega)$ can be accurately extracted from the
simulation data for $F(q,\tau)$. The static structure factor, Fig.
\ref{fig:sq}, and the static response, Fig. \ref{fig:chiq}, compare
very favorably with the experimental results, the discrepancy visible
in $S(q)$ at the smallest value of $q$ being due to the finite
temperature at which the measurements are performed. The $f$--sum
rule, $\partial F(q,\tau)/\partial\tau|_{\tau=0} = \int d\omega
~S(q,\omega) \omega = q^2$, is also verified with high precision.

\begin{figure} \epsfxsize=12cm \begin{center} \leavevmode
\epsffile{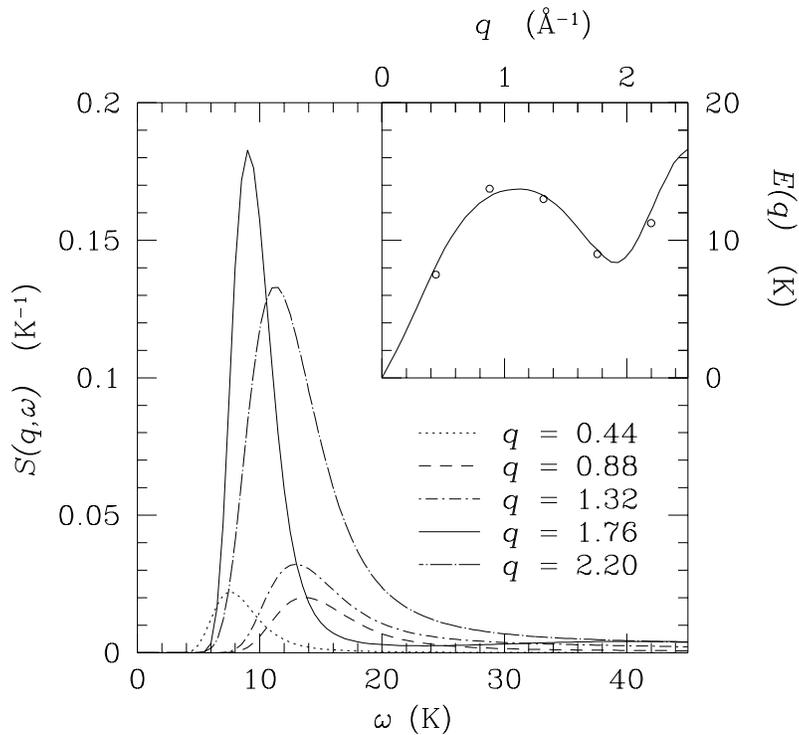} \end{center} \caption[]{Maximum Entropy
reconstruction of the dynamical structure factor. In the inset the
position of the maxima of the calculated $S(q,\omega)$ is compared
with the experimental excitation spectrum \cite{donnelly}.}
\label{fig:roton} \end{figure}

Inferring dynamical properties from imaginary-time correlations, on
the other hand, is much harder. Since the inverse Laplace transform
\eqref{sqom} with incomplete and noisy data for $F(q,\tau)$ is an
ill--conditioned problem \cite{maxent}, a least $\chi^2$ approach to
pin down a parametrized form for $S(q,\omega)$ is doomed to
failure. Additional constraints can be set on the solution
$S(q,\omega)$ by resorting to Maximum Entropy (ME) methods
\cite{maxent}.  We follow the implementation used in
Ref. \cite{massimo} to process data for $F(q,\tau)$ obtained at finite
temperature with a PIMC simulation.  The results are qualitatively
similar to those obtained in Ref.  \cite{massimo}. The ME
reconstruction of $S(q,\omega)$, shown in Fig. \ref{fig:roton}, is too
smooth and does not reproduce the sharp features exhibited by the
experimental structure factor. Furthermore, the relatively poor
quality of the available Monte Carlo data does not allow for a
reliable estimate of the statistical uncertainty on the results
\cite{maxent}. Nevertheless we do recover some useful information on
dynamical properties: the presence of a gap in the excitation spectrum
is unambiguously revealed, and the position of the peak of the
reconstructed dynamical response closely follows the experimental
dispersion of the elementary excitations (see the inset of figure 5).

We now outline the calculation of the superfluid density $\rho_s$.
Although the value of $\rho_s/\rho$ is trivially one for pure bulk
$^4$He in the ground state, interacting Bose systems in the presence
of an external disordered potential undergo a zero temperature
superfluid--insulator transition as the strength of the potential
increases \cite{dirty}.  We thus consider a model system of static
impurities in $^4$He.  The external potential $V_{ext}$ is represented
by attractive Gaussians, $V_{ext}({\bf r})=\sum_j A \exp[-\alpha ({\bf
r}-{\bf R}_j)^2]$, where the positions ${\bf R}_j$ of the impurities
are placed randomly in the simulation box, $A=-50\rm ~K$ and $\alpha =
0.5$ \AA$^{-2}$.  The trial function is multiplied by a one--body
factor $\exp[-\sum_{i,j}f(|{\bf r}_i-{\bf R}_j|)]$ which tends to
localize the $^4$He atoms around the impurities. No average over
different realizations of disorder was performed.

In a finite-temperature calculation with periodic boundary
conditions, $\rho_s$ can be estimated \cite{ceperley_rmp,ceperley} as
$\rho_s/\rho = \langle w^2 \rangle/(6\tau N_P)$, where the winding
number ${\bf w} = \sum_{i=1}^{N_P} \int_0^\tau d\tau' \left[
\frac{d{\bf r}_i(\tau')}{d\tau'} \right]$ is the displacement of the
center of mass of the system, and $\tau$ is the inverse
temperature. By taking the limit $\tau\to\infty$, appropriate for a
ground state calculation, the superfluid density at $T=0$ can be 
expressed in terms of the diffusion coefficient of the center of 
mass motion, $\rho_s/\rho = \lim_{\tau\to\infty} D(\tau)$, where 
$D$ is defined in Eq. \eqref{cm}.

\begin{figure} \epsfxsize=12cm \begin{center} \leavevmode
\epsffile{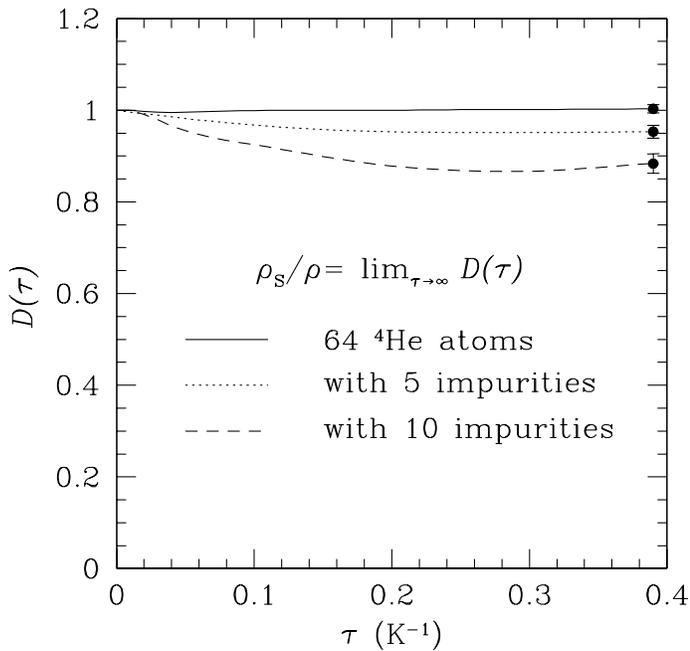} \end{center} \caption[]{The diffusion coefficient
of the center of mass motion.} 
\label{fig:rhos}
\end{figure}

The results reported in Fig. \ref{fig:rhos} show that the superfluid
fraction, which is correctly one for the pure system, is indeed reduced 
in the presence of the impurities. Longer simulations would be needed 
to relate the depletion of the superfluid fraction induced by the 
disordered potential to changes of the excitation spectrum.

\subsection{Comparison with other methods}

The RQMC method utilizes the Metropolis algorithm to sample an
explicitly known probability distribution, namely a discretized path
integral expansion of $\exp(-\tau H)\Phi_0$ which becomes exact in the
limit $\tau\to\infty$ and $\epsilon\to 0$.  Sampling a distribution,
as opposed to carrying weights, avoids the difficulties associated
with fluctuating weights which plague applications of `Pure Diffusion'
\cite{caffarel} or `Single Thread' \cite{nightingale_III} Monte Carlo.
For instance, we were unable to get converged results for our 64
particle system with PDMC.

The idea of sampling a path-integral representation of the
imaginary-time evolution to compute ground state properties is not new
\cite{nightingale_IV}. For instance, Variational Path Integral
\cite{ceperley_rmp} (VPI) uses the pair product approximation to
expand the many--body density matrix $\exp(-\tau H)$ and the bisection
method to sample the path, just like in the usual Path Integral Monte
Carlo method \cite{ceperley}.
 
RQMC features instead an expansion of the imaginary-time propagator
based on the Langevin dynamics generated by the trial function, and a
reptation algorithm to sample the exponential of the resulting
action. From the computational point of view, the advantage of this
particular choice can be understood in the limit of perfect importance
sampling: if the trial function is exact the local energy is a
constant, and reptation moves consisting of an arbitrary number of
time slices will be accepted with probability 1. Eventually, a very
poor wave function (or equivalently a very large number of particles)
will force us to take extremely small reptation moves, and moves of
the kind used in VPI will become more efficient \cite{ceperley}. VPI
has been implemented \cite{kevin} for the simulation of superfluid
$^4$He at $T=0$. According to the author of the VPI calculation, for
the system size considered here RQMC is considerably more efficient.

Sampling an explicitly known distribution is to be contrasted to the
standard BDMC, which samples an unknown distribution ({\it i.e.} the
mixed distribution $\Psi_0\Phi_0$) obtained from the asymptotic
solution of a differential equation \cite{mitas,nightingale_V}. BDMC
is designed to compute efficiently the total energy; however it
introduces a population-control bias, yields a mixed estimate for
operators not commuting with $H$, and does not retain direct
information on the imaginary-time correlations.  This information can
be retrieved through the `forward walking' technique
\cite{nightingale_V,kalos_fw,reynolds} and used to correct both the
population control and the mixed estimate biases, but at the price of
additional statistical fluctuations. We believe that RQMC will turn
out to be advantageous over BDMC in those cases where recovering
information from fluctuating weights becomes too noisy.

The results for the total energy shown in Table \ref{tab:table1} are
obtained with the RQMC and the BDMC methods using the same time step,
number of particles and trial function. From the estimated
statistical error we infer that BDMC is roughly 3 times faster than
RQMC for the calculation of the total energy.
Also listed in Table \ref{tab:table1} are the unbiased estimates for
the potential energy. In this case the BDMC algorithm with forward
walking (implemented in the ``backward storing mode'' described in
Ref. \cite{nightingale_V}) turns out to be roughly two times slower
than RQMC. Obviously, factors 2 or 3 for a couple of observables in a
particular physical system are not a conclusive assessment of the
relative performance of two algorithms. However the resulting factor
6 in the relative improvement of RQMC when BDMC has to be complemented
with forward walking suggests that, whenever explicit information on
imaginary-time correlations is used, RQMC is likely to be competitive
or better.

\section{Conclusions}

The most attractive feature of the RQMC method is that it uses the
dynamical properties of the random walk in a way that is directly
related to the imaginary-time properties of the physical system. A
previous implementations of similar ideas (PDMC, \cite{caffarel}) was
based on re-weighting and the scope of its applications was severely
restricted by the fluctuations of the weights. We have shown that, by
simply complementing the PDMC method with a re-sampling of the paths
based on the value of the action, the resulting RQMC algorithm can
afford system sizes typical of current BDMC simulations of continuous
strongly interacting systems. Unbiased estimates, static responses and
some insight into dynamical properties can be readily obtained. The
dependence of the computational effort on the number of particles and
on the quality of the trial function remains to be investigated. Such
an analysis will determine whether this method can be useful in more
general situations than $^4$He bulk liquid. Clusters, films and
superfluids in restricted geometries are natural candidates for
further applications.

For Fermion problems one has either to resort to the fixed-node
approximation \cite{mitas}, or to cope with the sign problem
\cite{kalos}. In the former case, the dynamical information contained
in the path is incorrect \cite{caffarel}, but still the explicit
knowledge of the weights along the path makes the algorithm free from
the mixed distribution and the population control biases; furthermore
it gives easily access to interesting quantities, such as a
low--variance estimator of Born--Oppenheimer forces in electronic
systems \cite{zong}. In the latter case the dynamics is correct, and
can be used for example to get information on the ground and excited
states from the imaginary-time evolution in the transient regime
\cite{bernu}; in similar cases however the real bottleneck will remain
the sign problem.

RQMC, VPI \cite{ceperley_rmp} and the technique discussed in Ref.
\cite{nightingale_IV} sample explicit expressions of the
imaginary-time propagator with the Metropolis algorithm to calculate
zero temperature properties. We believe that these methods are very
promising and deserve more attention than they have received so far.
All these methods are based on the Metropolis algorithm: they enjoy
therefore of a large freedom in the choice of the transition
probability \cite{umrigar}, which is probably not yet fully exploited.

\acknowledgements We are indebted with Kevin Schmidt for useful
discussions and for communicating to us unpublished
details of his VPI calculations on $^4$He. We are grateful to Matteo
Calandra for lending himself to be the first reader of these lecture
notes and for helping us improve them as much as we could.


\begin{thebibliography}{99}
\bibitem{ergodicity} This statement is only valid if $v(x) < +\infty$
everywhere inside the domain of $x$.
\end{thebibliography}
\end{document}